# Improving Spectral Resolution via Simultaneous Extracochlear and Apical Ground Contacts: A Neural Simulation Study

Jared A. Rybarczyk,[1] Erin L. Bratu, Ph.D.[2], Robert F. Labadie, M.D. Ph.D.[3], Jack H. Noble, Ph.D.[1]

[1] Department of Electrical and Computer Engineering – Vanderbilt University

[2] University of Rennes, INSERM LTSI - UMR 1099, 35000, Rennes, France

[3] Department of Otolaryngology – Medical University of South Carolina

ALL CORRESPONDENCE SHOULD BE ADDRESSED TO: Jared A. Rybarczyk, email: jared.a.rybarczyk@vanderbilt.edu, address: 2301 Vanderbilt Place Nashville, TN 37235

FINANCIAL DISCLOSURES/CONFLICTS OF INTEREST: This research was funded by NIH grant number R01DC014037 and R01DC022099. The authors have no relevant disclosures or conflicts to report.

## ABSTRACT

Objectives: While cochlear implants (CI) can provide excellent speech perception outcomes, speech recognition in noise remains a consistent problem for many patients due to channel interaction and spectral smearing. Previous attempts to address spectral smearing, such as current focusing stimulation techniques, signal interleaving, or sound coding strategies can help mitigate the channel interactions that lead to poor spectral resolution. Recently, works by the Landsberger group showing the clinical feasibility of an apical return contact placed in the cochlear helicotrema have presented new possibilities in terms of experimental stimulation schemes. This study explores the interactions between apical and extracochlear ground contacts, aiming to characterize the effects of dual ground usage on auditory nerve fiber (ANF) dynamics and determine their utility in improving spectral resolution.

Design: 6 patients with CIs were used in this retrospective study. Volume conduction models encompassing patient specific electro-anatomical characteristics and ANF compartment models defining neural dynamics were used to generate activation thresholds for 75 evenly spaced fiber bundles throughout each modeled cochlea. For each patient, 4 models were defined: an extracochlear ground model (corresponding to the typical CI configuration), an apical ground model (corresponding to a reference lead placed in the cochlear apex) and two combined ground models (corresponding to simultaneous usage of apical and extracochlear grounds with either equivalent voltage or equivalent current flow imposed). We compare our intra-patient models in terms of threshold shifts and standard deviation of fiber proportions to determine the effects of ground configuration used on channel interaction and spectral smearing.

Results: For all 6 patients, we observe shifts in ANF activation thresholds for the combined ground models indicative of more specific fiber stimulation per contact. Additionally, our combined ground models exhibit reduced fiber proportion standard deviation for several contacts of the stimulating array when compared to the standard extracochlear ground configuration or apical ground configuration.

Conclusions: The shifts in activation thresholds and reduction in standard deviations suggest that stimulation schemes utilizing extracochlear and apical grounds in conjunction can provide better spatial specificity and enhanced spectral resolution for certain contacts in an array. For patients with implanted apical grounds, stimulation schemes that make use of solely extracochlear or apical grounds in addition to combined schemes could provide the best speech outcomes.

## INTRODUCTION

Cochlear Implants (CI) are state-of-the-art neuro-prostheses that aim to partially restore hearing in patients with mild-to-profound sensorineural hearing loss. For the nearly 1 million people worldwide who have undergone CI surgery, post-operative hearing outcomes are generally positive (McRackan et al., 2018), with one meta-analysis finding average word perception increasing by ~46% and self-reported benefit increasing by ~22% (Boisvert et al., 2020). One aspect CI patients frequently struggle with following implantation is understanding speech in environments with masking noise, creating difficulty in obtaining reasonable speech perception in many everyday situations (Friesen et al., 2001). It is widely theorized that difficulty with speech perception and variability of hearing outcomes in the presence of masking noise is the result of spectral smearing (Fu & Nogaki, 2005; van de Velde et al., 2015; O'Neill et al., 2019). Typical CIs range from 12-22 electrodes per intracochlear array (Dhanasingh & Jolly, 2017), delivering stimulating pulses to auditory nerve fiber (ANF) cell bodies present within the spiral ganglion of the cochlear modiolus. These ANFs are arranged tonotopically throughout the cochlea in descending order, with more basal fibers corresponding to higher frequency sound and more apical fibers corresponding to lower frequencies. Due to the large number of ANFs present in the cochlea (estimated between 30,000 and 37,500 for an average human (Nadol, 1988; Siegel, 2009)), stimulation from the comparatively sparse number of electrodes in a typical CI can result in poor tonotopic alignment and spatial frequency overlap, causing issues such as channel interaction. Channel interaction specifically occurs when there is overlap between the populations of ANFs activated by distinct contacts, resulting in poorer hearing outcomes and spectral smearing (Goehring et al., 2021).

As the detrimental effects of masking noise on speech perception in CI users are both common and variable in presentation and severity, many different strategies have previously been developed to address them. Previous studies have shown that channel interaction due to multiple simultaneously active electrodes are a function of contact spacing, individual cochlear anatomy, and stimulation geometry e.g. monopolar or bipolar stimulation patterns (White et al., 1984). As such, some straightforward approaches to enhancing spectral resolution involve current shaping and the generation of narrower electric fields via multipolar stimulation. Prior studies attempting current shaping and focusing have found varied degrees of benefit to spectral resolution, typically observing that current steering with tripolar stimulation offers better spectral channel discrimination and speech perception in noise than monopolar stimulation (Berenstein et al., 2008; Landsberger & Srinivasan, 2009; Srinivasan et al., 2013). Current focusing strategies can be especially useful in patients with large distances between contact and ANF cell bodies (DeVries & Arenberg, 2018). However, the benefits to spectral resolution are often minor, and the increased current required for tripolar configurations can limit its clinical feasibility (Zhu et al., 2012). Additionally, some studies have observed increased thresholds and threshold variability in partial tripolar configurations when compared to monopolar or bipolar configurations, which correlated with worsened speech outcomes (Bierer, 2007).

In addition to current steering techniques, sound coding strategies have also been explored in the context of reducing spectral smearing. Channel selection strategies such as Continuous Interleaved Sampling (CIS) and high rate N-of-M can potentially improve spectral resolution by selecting non-overlapping or dominant frequency bands and reducing the number of active channels per stimulus cycle (Wilson et al., 1993; Buechner et al., 2009). For users with bilateral implants, bilaterally interleaved processing strategies can offer improved spectral resolution by spreading the required stimulation across each ear. This can result in reductions in channel interaction while still providing the same total spectral information, but may require perceptual alignment to retain sound localization ability (Aronoff et al., 2016). More recent strategies such as Temporal Integrator Processing (TIPS), which focuses on removing stimuli that are perceptually minor, have found improvements in perception quality of up to 2.4dB with 50% of pulses removed (Lamping et al., 2020). Evaluations of TIPS found that while quality ratings of speech perception in noise were lower than CIS, it still outperformed the Advanced Combination Encoder strategy while reducing power consumption by up to 42% (Shahidi et al., 2026).

While bipolar and tripolar current steering strategies and speech coding techniques are the dominant methods for addressing common spectral resolution issues in CI recipients, studies examining different grounding techniques for the same purpose are relatively non-existent, remaining focused on managing stimuli related abnormal sensation or shifting stimuli altogether. Commonly, monopolar stimulation from intracochlear electrodes flows to extracochlear references, typically a distant lead or a plate implanted within the CI case. In some cases, particularly for patients with facial nerve stimulation (FNS), a technique known as multi-mode grounding (MMG) is applied (Zellhuber et al., 2022; Ordóñez Ordóñez et al., 2025). In addition to the typical extracochlear reference electrode, MMG schemes allow for every non-stimulating electrode in the array to act as a return path for current, effectively reducing current flow from the cochlea and reducing FNS. Similarly, the usage of apical ground contacts (placed in the helicotrema), has begun to be examined. Two main studies have clinically implemented apical ground leads for the purpose of improving apical ANF stimulation without a longer intracochlear array finding that the presence of an apical ground both apically biased current flow in cadaveric specimens and caused a perceived lower pitch without significantly affecting speech perception (Landsberger et al., 2022, 2024). These results suggest potential clinical feasibility and utility for future stimulation strategies.

While studies on novel ground configurations focus largely on FNS reduction and more apically directed current flow, the full potential of experimental stimulation schemes that make use of these grounds for reducing spectral smearing and enhancing speech perception in noise is currently unknown. This study aims to address that gap by implementing computational models of the cochlea and ANF to determine the effects of 3 different grounding schemes (extracochlear case ground, apical ground, and a novel dual ground) on neural dynamics and computationally derived spectral smearing metrics.

## MATERIALS AND METHODS

This is a retrospective study of 6 unilaterally implanted patients (3 with modiolar hugging arrays and 3 with lateral wall arrays, referred to as MH1-3 and LW1-3 respectively) that has received IRB approval. All computations presented were performed in either MATLAB R2025a or Python 3.11 (Python Software Foundation, 2022; The Mathworks Inc., 2022). Additionally, neural activation simulations and cable compartment models were created via the NEURON library (Hines et al., 2009).

## Electro-anatomical Modeling

For each of the 6 patients in this study, a two-part electro-anatomical modeling pipeline as described in (Bratu et al., 2026; Rybarczyk et al., 2026) was implemented. Patient post-operative CT scans were used to create volumetric anatomical maps, representing the CI array and 4 different tissue classes (perilymph/endolymph, soft tissue, neural tissue, and bone). By assigning realistic resistivity values to each of these surfaces and modeling 1250µA current sources at CI array contact locations, a solution to Poisson's Equation can be approximated via the Finite Difference Method (FDM). This produces a volumetric electro-anatomical voltage map, defining electrical potential values at each voxel of the solution domain due to stimulation from an arbitrary contact.

The FDM voltage map is then sampled along 70-85 points at each of 75 evenly spaced ANF bundles (each representing 500 individual fibers to model a total 37,500 ANFs per ear), and these sampled voltages are used as inputs to ANF cable compartment models. Specifically, this study utilizes the Briaire and Frijns ANF model (Briaire & Frijns, 2005), supplying the sampled voltage values as singular cathodic-leading biphasic extracellular stimuli at either peripheral Nodes of Ranvier or central internodes. Using an optimization routine, the minimum threshold current from an arbitrary CI contact for 50% of a given ANF bundle to propagate an action potential can be determined. Activation thresholds are limited to a maximum current of 5000µA, well above the typical stimulus current used in everyday CI programming.

## Ground Schemes

In this study, 4 variations of the ground used in the FDM electro-anatomical model are applied to each patient. For the first scheme, the ground conditions are defined to be the edges of the solution domain, approximating the distant extracochlear ground (EG) used in most standard CI programming. The second scheme models a cylindrical apical ground (AG), approximately 0.75mm in diameter and 5mm in length inserted into the fluid of the cochlear apex as described in (Landsberger et al., 2024). The third scheme (combined grounds with equivalent voltage, EqV) combines both the EG and AG simultaneously, forcing both to act as ground conditions for a given FDM simulation and representing holding both grounds at equivalent 0 voltage. The final scheme (equivalent current, EqI) also simulates the use of combined grounds. However, instead of holding them simultaneously at 0 volts, the EqI scheme models equivalent current flow to each ground. To compute the EqI scheme, the ratios of voltage at each stimulating contact between the EG and AG voltage maps are used, such that for stimulating contact N:

Eq.1
$$EqI = 0.5EG + 0.5*(EG_N/AG_N)*AG$$

where $EG_N$ and $AG_N$ are the electrical potential values at stimulating contact N for the extracochlear and apical simulations respectively. Each of these four ground schemes are applied independently to the computational ANF model in order to determine respective neural stimulation thresholds. Figure 1 shows the relative locations of the modeled EG and AG scheme grounds to the patient cochlea.

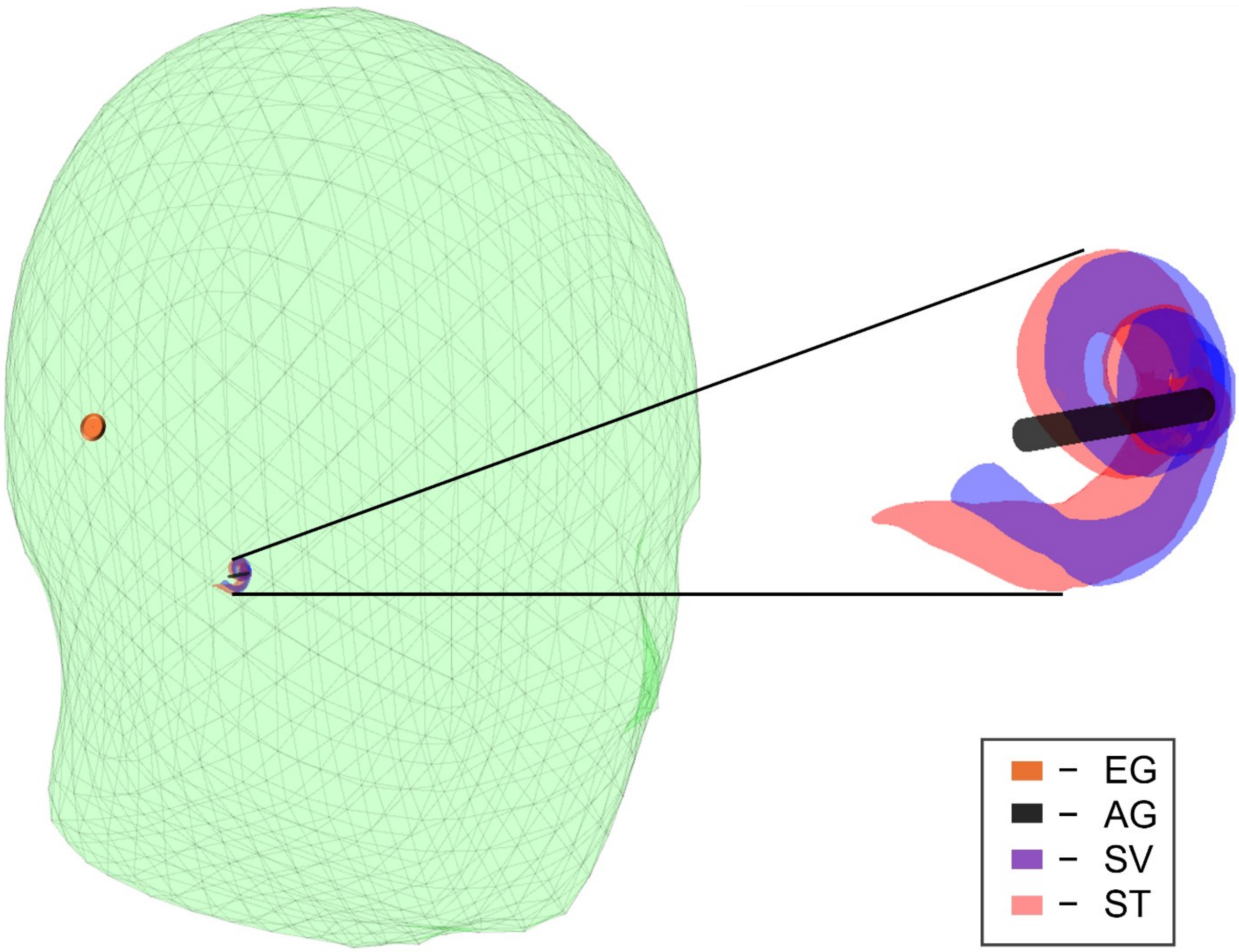

Fig. 1: Locations of the EG (orange) and AG (black) scheme grounds relative to the patient Scala Vestibuli (SV, blue) and Scala Tympani (ST, red). The EG represents the extracochlear ground and the AG represents the apical ground inserted into the perilymph of the cochlear apex. Scalp mesh generated from the Freesurfer software (Fischl, 2012).

**ANF Bundle Activation Proportions**

Modeling the broad behavior of ANFs within a given bundle allows for computation of the threshold value at which 50% of all ANFs in the bundle can propagate stimulation. It is known that the spiking probability of a single ANF with respect to stimulation amplitude can be represented with a sigmoid function (Rattay & Tanzer, 2022). As such, the spiking probability of all ANFs of a given bundle also follows a normal distribution with stimulus current. From this relationship, the threshold for a given bundle and the stimulus current from a given contact can be fit to a sigmoid function to produce the proportion of individual ANFs within the bundle such that:

Eq.2 $P_{XN} = \text{normCDF}((Y_N - \text{threshold}_x)/(0.25*\text{threshold}_x))$

where $\text{threshold}_x$ is the ANF model derived activation threshold for fiber bundle X, $Y_N$ is the stimulus current produced from contact N, and $P_{XN}$ is the proportion of fibers active in X due to stimulation from N. In order to facilitate worthwhile comparison between the 4 ground variants defined above, Eq.2 is used to determine adjusted stimulus levels for each contact in the AG, EqV, and EqI models relative to the EG model. Specifically, for each contact, Eq.2 allows for iterative adjustment of stimulus such that the total number of active ANF fibers per contact are equalized between all ground variants. For example, if the total number of active fibers across all 75 bundles due to stimulation from contact 1 in the EG model is 35,000, the stimulus current from contact 1 in the AG, EqV, and EqI models is adjusted until they reach 35,000 active fibers. It is important to note that this method only equalizes total active fiber count and does not effect the locations of the active fibers, i.e. the 35,000 active fibers of the EG model are almost certainly not identically located to the 35,000 fibers of the AG model. This equalization becomes necessary because the stimulus current defined in the electro-anatomical model is held constant between ground variants, meaning that for the same current, greater proportions of neural activation could be observed in EG schemes than AG schemes due to the difference in relative current flow. If the equalization was not performed, comparisons between ground schemes would not be valid since they would not represent identical levels of stimulation.

**Spectral Resolution Metrics**

From the active fiber proportions of each bundle, the mean and standard deviation (SD) of the fiber bundles can be calculated to better quantify the effects of each ground scheme on spectral smearing. For a given contact N, multiplying the active fiber proportion by the total number of fibers present in the bundle (500) produces the number of active fibers for a given bundle. For the active fiber proportions of all bundles $ANF_1$ through $ANF_{75}$ stimulated by N, taking the mean produces a value in the range of 1-75 corresponding to the location of the average ANF bundle stimulated by N (where 1 represents the most basal ANF and 75 represents the most apical ANF). Similarly, taking the SD produces a value corresponding to how spread-out stimulation from the mean ANF is, representing spectral smearing. In an ideal stimulation scheme the mean ANF would increase linearly with the stimulating contact while the SD would be 0, indicating only fibers of a single spatially matched ANF bundle were activated.

Prior to SD and mean by ANF bundle proportion calculations, active ANF counts in the AG, EqV, and EqI models were equalized with respect to the EG model to allow for valid comparison. The reference stimulus current ($Y_n$ in Eq. 2) for each contact in the EG model was set to 500µA. For all analyses, ANF bundles with less than 1% of fibers (less than 5 fibers) activated were treated as ANF bundles with 0 fibers activated to reduce the effect of noisy activations that would likely have imperceptible effect on spectral resolution outcomes.

## RESULTS

For each patient, the CI array contacts range from 1 (most basal) to 22 (most apical) and the ANF bundles range from 1 (most basal) to 75 (most apical). Figure 2 shows the computational ANF model derived neural activation thresholds (in μA).

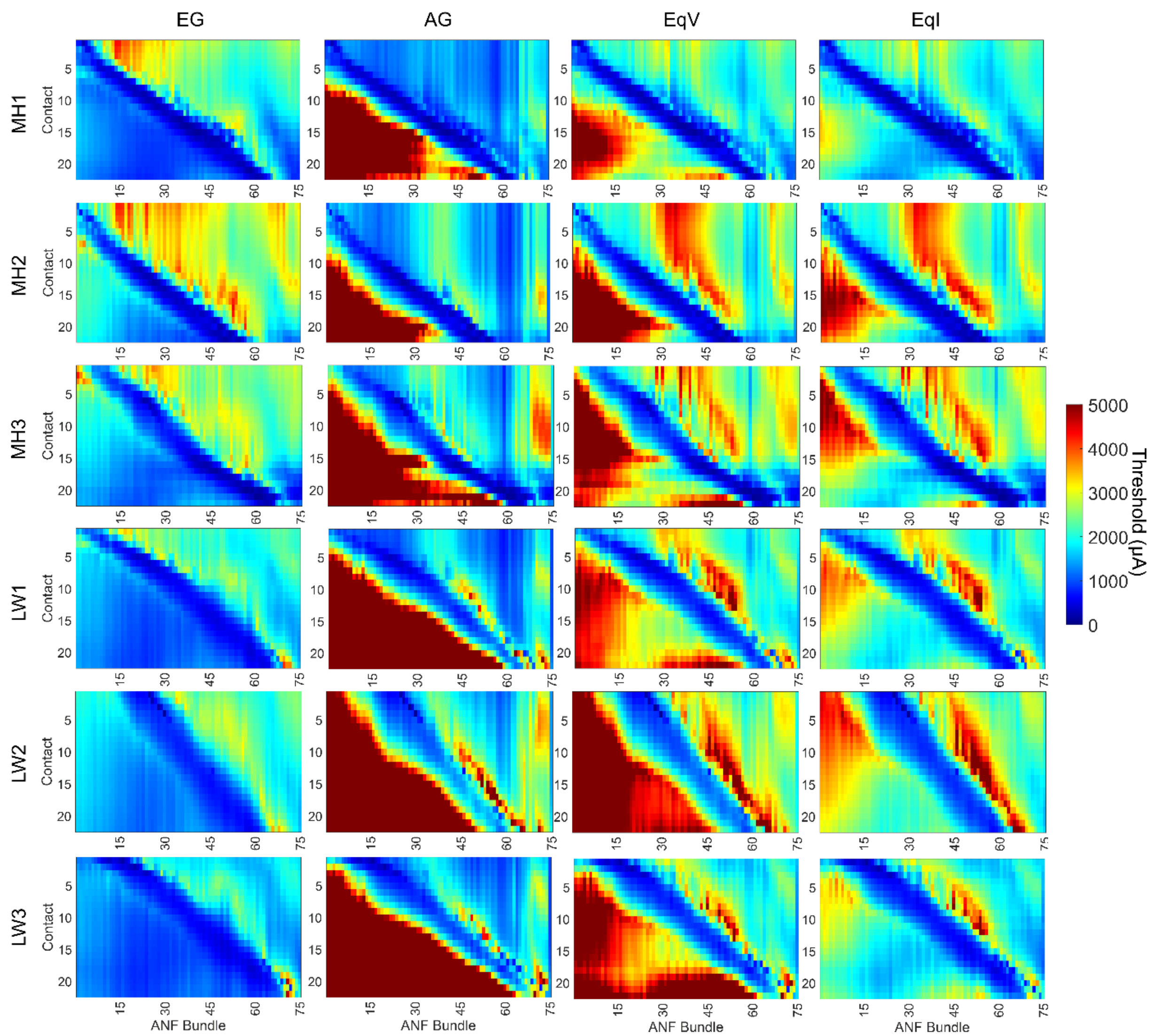

Fig. 2: Neural stimulation thresholds for patients based on each of the 4 ground stimulation schemes.

### Neural Activation Thresholds

In describing the distribution of neural thresholds throughout the different ground schemes, one landmark useful for comparison is the spatially matched low threshold (SMLT) region corresponding to areas where ANFs have low stimulus thresholds as a result of being nearest the stimulating contact. As this region is evident to varying degrees across all schemes tested, it is helpful to refer to the differences between ground schemes as regions that are more apical (to the right of) or more basal (to the left of) than the SMLT region.

For all 6 patients, when the EG stimulation scheme was applied, 2 main effects on neural activation thresholds were observed. First, for all 6 patients, EG scheme stimulation results in comparatively reduced basal ANF bundle activation thresholds, indicating that for all contacts in the array, the usage of an EG scheme promotes stimulation of more basal ANF bundles. Conversely, ANFs that are relatively apical to a given contact exhibit higher thresholds than the ANFs that are basal to the SMLT region.

In comparison to the thresholds observed in the EG scheme, the AG stimulation scheme exhibits an inversion of threshold magnitude trends. Specifically, for all 6 patients, the ANFs that are basal to the SMLT region exhibit higher activation thresholds than the ANFs that are more apical to the SMLT region. This indicates that, in the AG scheme, ANF bundles that are more apical than a given contact are easier to stimulate than ANF bundles that are more basal, the opposite of the relative threshold distribution observed in the EG scheme.

As the stimulation thresholds generated by the EG and AG stimulation schemes are comparatively opposite in the locations of equivalent magnitudes, when the two schemes are combined in the EqV and EqI stimulation schemes threshold distributions reflecting this combination are observed. Specifically, for the EqV scheme, the SMLT region is accompanied by comparatively higher thresholds for both relatively apical and basal ANFs, with the highest thresholds being observed in the relatively basal ANFs for all 6 patients.

For all 6 patients, in comparison to the EqV stimulation scheme, the EqI stimulation scheme exhibits similar trends in threshold distribution with slightly reduced contrast between the SMLT region and regions that are more apical or basal. Specifically, this reduction in contrast corresponds to the ANF bundles spatially basal relative to a stimulating contact. However, while the contrast between the SMLT region and relatively basal region is reduced, the thresholds of the ANFs relatively apical to the SMLT region remain essentially identical between the EqV and EqI schemes. The magnitude of thresholds along the SMLT region are similarly unchanged between the EqV and EqI schemes.

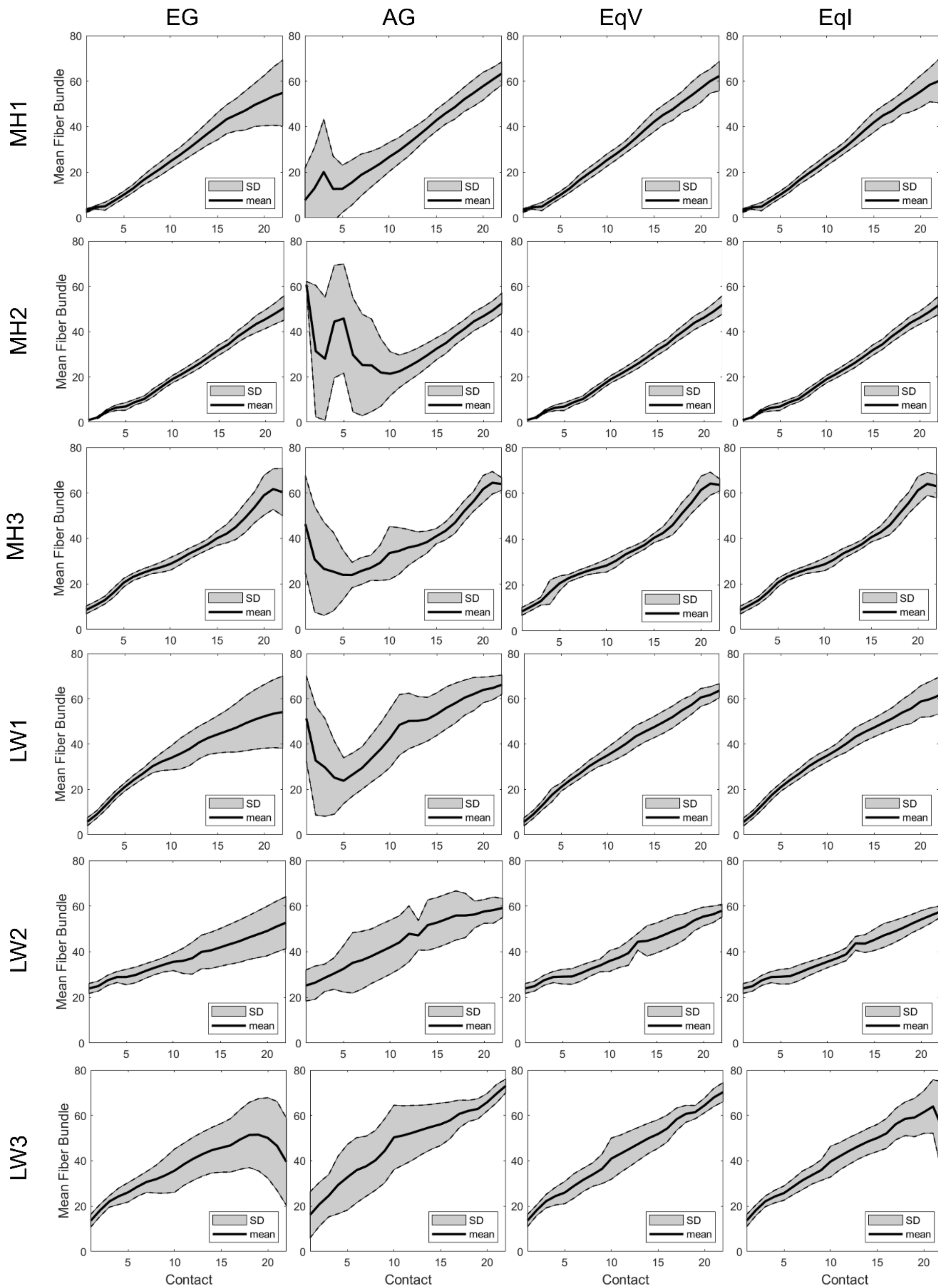

Fig. 3: The mean and SD for each contact in patient CI arrays based on each of the 4 ground stimulation schemes.

**Standard Deviation and Mean by Fiber Number**

Figure 3 shows the mean active fiber and SD due to stimulation in each of 4 ground schemes. Between all 6 patients, clear trends for the 4 different stimulation schemes emerge. Specifically, the EG scheme exhibits low SD values for basal contacts, while apical contacts exhibit comparatively increased SD. This behavior corresponds to the bottlenecked shapes apparent in Fig. 3, with the SD tending to increase between contacts 3 (for patient LW3) and 15 (for patient MH2), representing where reductions in spectral resolution in the EG scheme would be expected. With respect to the mean, the EG scheme plots tend to follow a roughly positive linear correlation with contact. An exception comes in the form of patients MH3 and LW3, where the peak mean value and the mean value at the final contact are not identical. This drop-off in mean could be related to anatomical variations and the helical nature of the cochlea, such that the deepest contacts recruit more relatively basal fibers than their neighbors due to wrap-around.

For the AG scheme, observed SDs and means were radically different when compared to the EG scheme. Prominent jumps in the mean active ANF bundle were observed for 4 of the 6 patients, and this characteristic was more evident for spatially basal contacts, resulting in a "U" shape, where the most basally located mean bundle was stimulated by contacts towards the middle of the implant array. The highest fluctuations in mean active bundle were observed in patient MH2, with the differences between contacts 1 and 2 being ~30, indicating a large spatial shift in stimulation. However, patients LW2 and LW3 exhibited more linearly correlated mean bundles, similar to the generally observed EG scheme trends. These jumps in mean active bundle were accompanied by large SDs in all 6 patients. However, for more apical contacts, there was a marked reduction in SD. In the case of patient MH1, for example, the AG scheme has a lower SD for contacts >15 than the EG scheme.

For the EqV and EqI schemes, there are overall reductions in SD for all patients when compared to EG and AG schemes. For patient MH1, both the EqV and EqI schemes have lower SDs for more apical contacts as compared to the EG scheme and basal contacts as compared to the AG scheme. The EqV scheme particularly observes lower SDs in the apical contacts in comparison to the EqI scheme. Additionally, the mean active bundle increases at a higher rate with respect to contact for the EqV and EqI than the EG scheme. Similar trends are observed in patient LW1, with EqV and EqI schemes showing steeper increases in mean active fiber and lower SDs in both basal and apical regions compared to AG and EG schemes respectively. For patient MH2, the differences between the EG and EqV/EqI schemes are less pronounced. While the EqV and EqI schemes both show reduced SD in apical contacts, it is to a lower extent than that observed in MH1 and LW1. Additionally, the slope of the mean is roughly equivalent across all contacts for the 3 methods. The contrast in apical contact SD and means between the EqV/EqI and AG schemes are similar to the trends observed in the other patients.

For the EqV and EqI, all 6 patients exhibit general reductions in SD across both apical and basal regions when compared to the EG or AG schemes alone. This result mirrors the concept of combination observed in the neural stimulation thresholds, wherein the lowest SDs for the EG scheme (typically basally located) and the lowest SDs for the AG scheme (typically apically located) combine to provide reduced SDs across each contact in the patient array. Additionally, the overall slope of the mean bundle is increased in the EqV and EqI cases when compared to the EG case, suggesting that these schemes stimulate a wider range of fiber bundles. Though both EqV and EqI schemes outperform the corresponding EG and AG schemes, differences between them emerge across the patients. When compared to the SD of the EqV scheme, the EqI results show a similar, yet slightly increased general SD for 5 of the 6 patients, indicating slightly reduced overall spectral resolution while still outperforming the EG or AG schemes alone. Only in patient LW2 was the SD for the EqI scheme lower in both basal and apical regions than the EqV scheme. Some further regional exceptions to the EqV scheme's outperformance can be observed, however, as in the cases of MH3 or LW3 where the SD in the EqI scheme around contacts 3-6 (for MH3) or 3-15 (for LW3) is lower than the corresponding value for the EqV scheme. Important to note, however, is that while in general the EqV scheme yields lower SD than EqI, with the exception of patient LW3 the mean bundle values between the two schemes are nearly identical. In the case of LW3, the mean values remain similar until the final 5 contacts in the array, where the EqI scheme exhibits reduction (similar to the EG scheme) while the EqV scheme remains positive in slope (similar to the AG scheme).

## DISCUSSION

The issue of low spectral resolution in CI users is one that has been extensively studied and addressed to varying degrees of success. In this study, 2 new methods of enhancing spectral resolution via combined usage of an extracochlear and apical ground were examined, and neural stimulation thresholds and spectral resolution metrics were calculated to examine the degree of improvement offered. For all patients it was observed that the EG stimulation scheme produced broadly reduced thresholds in regions that are basal with respect to a given contact. As the EG scheme is representative of the typical ground used in many CI users, the comparatively high thresholds in regions more apical than a given contact are indicative of the commonly observed difficulty in stimulating low frequency encoding ANFs. The AG scheme, which was experimentally devised to address this issue in (Landsberger et al., 2022), observed shifted thresholds such that ANFs more apical than a given contact were easier to stimulate and more basal ANFs had higher thresholds. Within each of these schemes, the ANF bundles with the lowest thresholds remained those that were spatially nearest the contacts in the SMLT regions. Thus, by combining the EG and AG schemes in the EqV and EqI schemes, the lowest thresholds remain in those spatially matched ANF bundles, while thresholds for bundles outside the SMLT region increased. These threshold shifts indicate potentially enhanced spectral resolution.

The potential increase in spectral resolution via the EqV and EqI schemes is further supported by the SD and mean fiber bundle results. The overall reductions in SD and the steeper slope of the mean active bundle observed in the EqV and EqI schemes further supports not only an improved specificity in ANF stimulation, but a higher degree of stimulus matching. In this way, the EqV and EqI schemes are capable of combining the strengths of both the EG and AG schemes, potentially providing narrower matched stimulus to basal and apical ANF bundles alike.

Though the differences between the EqV and EqI schemes were much less pronounced than the differences between either of them and the EG or AG schemes, they were observed nonetheless. The observed trend of the EqV schemes having lower SD than EqI schemes while retaining nearly identical mean bundle values indicates that though the spread of stimulation in the EqI scheme may be wider, it is centered at the same spatial location as the EqV scheme.

One important aspect to consider is the clinical feasibility of these stimulation techniques. Previous work has already shown that the placement of an apical ground contact into the cochlear helicotrema via the round window is not only possible, but possible without noticeable detriment to speech outcomes (Landsberger et al., 2022). As such, the simultaneous activation of extracochlear and apical grounds is seemingly possible. With respect to implementing the EqV or EqI schemes described above, the more clinically feasible method would seem to be the EqV scheme, as activating each ground present is a physically easier task than enforcing that equal amounts of current flow to two spatially different grounds.

Future directions based on this work should focus on verifying clinical utility and exploring more combinations of grounding schemes. As a computational study, this work serves to outline the potential utility of a novel CI stimulation scheme focused on improving spectral resolution. The potential benefits to reducing smearing and channel interaction are explored through the lens of neural dynamics and the effects of speech perception are extrapolated from this. However, qualitative results and speech recognition outcomes in real patients remain the gold standard for assessing the usefulness of any stimulation scheme. As such, emphasis should be placed on recruiting and implementing the methods of this study in real patients to verify the conclusions drawn.

Additionally, the creation of hybrid stimulation schemes that combine the EG, AG, and EqV or EqI schemes on a per-contact basis may be of interest in further improving spectral resolution. While the results of this study indicate that EqV or EqI schemes perform better than EG or AG schemes as a whole, on a contact-to-contact basis there are select instances where the EG scheme achieved, for example, lower SD than the combined schemes. This is particularly observed in patient MH2, where the EG scheme achieves a slightly lower SD for the first 5 contacts than either the EqV or EqI schemes. As such, it is possible that for some patients, selectively determining which contact uses which ground scheme may be advantageous in optimizing spatial specificity and improving spectral resolution. Additional studies on the usage of combined ground schemes with bipolar or tripolar stimulation as opposed to monopolar may also offer enhancements in spatial specificity.

**Conclusion**

The results of this study indicate the potential of stimulation schemes that utilize simultaneous apical and extracochlear grounds to improve spectral resolution beyond what can be achieved with either ground alone. It is possible that, by using both grounds in conjunction, stimulation can be made more selective, possibly reducing channel interaction and improving hearing outcomes for CI users in noise.